\documentclass[journal]{IEEEtran}
\usepackage{cite}

\ifCLASSINFOpdf
\else
\fi
\usepackage{amsmath}
\usepackage{algorithmic}

\usepackage{array}

\usepackage{graphicx}
\usepackage{fixltx2e}

\usepackage{stfloats}
\usepackage{textcomp}
\usepackage{epstopdf}
\usepackage{multirow}
\usepackage[table]{xcolor}
\usepackage[section]{placeins}
\usepackage{float}

\begin{document}
%
\title{Topology Identification in Distribution Networks using Harmonic Synchrophasor Measurements}
%
%
%

\author{Lei~Chen,~\IEEEmembership{Student Member,~IEEE},
        Mohammad~Farajollahi,~\IEEEmembership{Student Member,~IEEE},
        Mahdi Ghamkhari,~\IEEEmembership{Member,~IEEE},
        Wei~Zhao,
        Songling~Huang,~\IEEEmembership{Senior Member,~IEEE},
        and~Hamed~Mohsenian-Rad,~\IEEEmembership{Fellow,~IEEE}
\thanks{Lei Chen, Wei Zhao, and Songling Huang are with the State Key Lab. of Power System, Department of Electrical Engineering, Tsinghua University, Beijing, 100084, China.}
\thanks{Mohammad Farajollahi and Hamed Mohsenian-Rad are with the Department of Electrical and Computer Engineering,
University of California, Riverside, CA, USA.

Mahdi Ghamkhari is with the Electrical and Computer Engineering Department of University of Louisiana at Lafayette, Lafayette, LA, USA.
}
}

%
%

\markboth{}%
{Shell \MakeLowercase{\textit{et al.}}:}
%



\maketitle

\begin{abstract}
Topology identification (TI) in distribution networks is a challenging task due to the limited measurement resources and therefore the inevitable need to use pseudo-measurements that are often inaccurate.
To address this issue, a new method is proposed in this paper to integrate \emph{harmonic synchrophasors} into the TI problem in order to enhance TI accuracy in distribution networks.
In this method, topology identification is done \emph{jointly} based on both fundamental synchrophasor measurements and harmonic synchrophasor measurements.
This is done by formulating and then solving a mixed-integer linear programming (MILP) problem.
Furthermore, an analysis is provided to capture the number of and the location of harmonic sources and sensors that are needed to {ensure} full observability.
The benefits of the proposed TI scheme are compared against those of the traditional scheme that utilizes only the fundamental measurements.
Finally, through numerical simulations on the IEEE 33-Bus power system, it is shown that the proposed scheme is considerably accurate compared to the traditional scheme in topology identification.
\end{abstract}


%
\IEEEpeerreviewmaketitle

\section{Introduction}\label{introduction}
\IEEEPARstart{K}{NOWING} the topology of distribution network (DN) is crucial for power distribution system operation, and comes with various applications such as event source location \cite{Mohammad2018}, state estimation \cite{DSSE2017}, and line impedance estimation \cite{Paolo2019}.
The topology is identified if one knows the status of the switches in all distribution line segments.
If the changes in the status of the switches are not identified, the accurate topology of the power system will be lost and large errors will occur in the mentioned applications that rely on knowing the topology of the DN.

The problem of topology identification is of importance in both distribution and transmission networks.
In transmission networks, there are often either sensors that directly identify the status of switches or there are sufficient measurements to achieve full observability to estimate the status of switches.
Several methods have been previously proposed to address topology identification in transmission systems, such as in \cite{HM1999,GSE1998,GSE2004,GSE2015,Joint2012}.


However, conducting topology identification is more challenging when it comes to distribution systems.
In particular, due to the lack of sufficient measurements in DNs, there is often a need to highly rely on pseudo-measurements  in order to solve the TI problem.
For example, pseudo power injections \cite{Mohammad2019,Joint2020,MIQP2016} and pseudo current injections \cite{Abur1995} have been previously used.
However, given the inherent inaccuracy in pseudo-measurements, they may cause incorrect topology identification.

In this paper, we propose to utilize harmonic measurements to resolve the above issue.
This idea is motivated by the fact that, harmonic currents are ubiquitous in DNs as a result of widespread utilizations of non-linear devices such as power electronic inverters.
Also, a new generation of phasor measurement units (PMUs) have been developed recently that can report not only the fundamental synchrophasors but also the harmonic current synchrophasors with high {accuracy} \cite{GPS2009,Carlo2009}.
PMUs of this generation are already installed in multiple pilot utilities, such as the ones in Japan \cite{Japan2003}.
The idea of using harmonic currents in a TI scheme is specially reinforced by the fact that a group of researchers has developed various harmonic synchrophasor estimators \cite{Lei2018,HPE2019,Extended2014} which come with applications in harmonic state estimation \cite{MeloHarmonic,MELO2019303} and high impedance fault location \cite{Hamed2017}.

One outstanding merit of the proposed TI scheme is that it does \emph{not} require placement of additional PMUs in DNs but rather utilizes \emph{additional information}, i.e. harmonic currents, available from the existing PMUs.

The proposed study is also motivated by the fact that, incorporating harmonic currents in TI is not a trivial task.
To see this, notice that there are two features in harmonic current phasors that make them quite different from fundamental current phasors: 1) the sources of harmonic currents are located in the load side in contrast to the source of fundamental current that is in the substation side; and 2) only a few types of loads can generate considerable amounts of harmonic currents.

The main contributions of the paper are as follows:
\begin{itemize}
  \item To the best of our knowledge, this is the first paper that proposes using harmonic synchrophasor measurements to identify the topology of distribution system.
  \item The proposed method works by integrating a harmonic TI problem formulation with a fundamental TI problem formulation through introducing several conjunction equations to leverage both fundamental and harmonic measurements to identify the distribution network topology.
  \item The problem is originally formulated in form of a nonlinear mixed integer program. Several linearization steps are developed to convert the problem into a tractable mixed integer linear program (MILP).
  \item Through computer simulations, it is shown that the accuracy of the proposed TI scheme is at least 10$\%$ more than {that of the traditional scheme which} uses only measurements of the fundamental currents.
  \item An analysis is provided that specifies the number of and the location of the harmonic resources and PMUs that are required to {ensure full observability of the DN.}
\end{itemize}
This paper gives focus to the study of TI in radial networks, as DNs are mostly with radial topology \cite{Abur1995}.

\begin{figure}
  \centering
  \includegraphics[width=8cm]{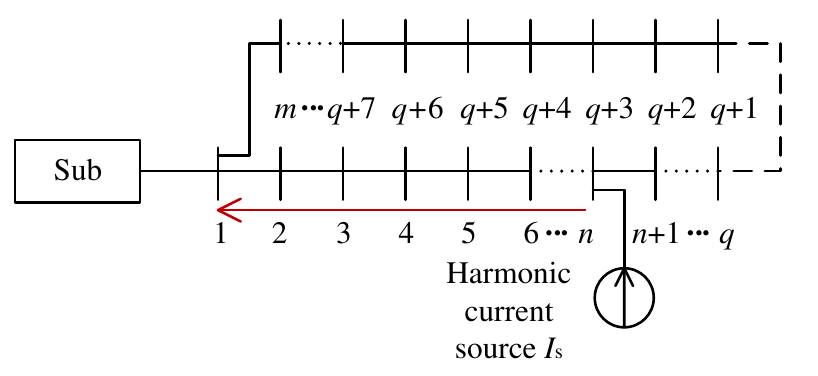}\\
  \caption{An illustration of harmonic current flow in radial DNs. The branch ($q$, $q$+1) is normally open, and other branches are normally closed. The harmonic current path is shown {by a} red arrow line, i.e., from node $n$ to the substation.}\label{fig_Sample_network}
\end{figure}

\section{Harmonic Currents in Radial Networks}\label{sec_analysis}

According to the Norton Theorem and concerning the harmonic currents, the nonlinear loads can be modeled by a harmonic current source and a shunt admittance \cite{DSS}.
We assume that there exist no harmonic resonance in the DN under study.
This assumption is justified by the fact that, harmonic resonance may {occur} only when the resonant frequency coincides with a harmonic frequency, which {may happen only rarely}.
Besides, there exist mature techniques to damp harmonic resonance \cite{Akagi2000}.
Under the said assumption, the load shunt impedance is much larger than the equivalent harmonic impedance of the substation, see \cite{DSS} for more details.
As a result, almost all of the harmonic source current is injected into the substation and the harmonic current injected to the load can be neglected.

\begin{figure}
  \centering
  \includegraphics[width=7cm]{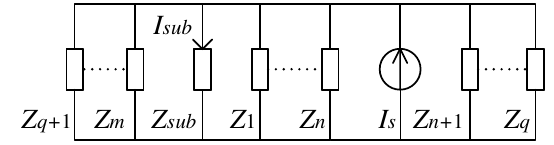}\\
  \caption{Equivalent impedance network of the DN in Fig. \ref{fig_Sample_network}. $Z_{sub}$ is the equivalent impedance of the substation. $Z_{n}$ is the impedance of the load at node $n$.}\label{fig_impedance}
\end{figure}

Consider the power distribution system in Fig. \ref{fig_Sample_network}.
Its corresponding equivalent impedance network is as shown in Fig. \ref{fig_impedance}.
The harmonic current injection to the substation can be calculated as follows:
\begin{equation}\label{eq_current_injection}
\begin{aligned}
I_{sub}=\frac{\frac{1}{Z_{sub}}}{\sum_{n=1}^m\frac{1}{Z_{n}}+\frac{1}{Z_{sub}}}I_s
=\frac{1}{\sum_{n=1}^m\frac{Z_{sub}}{Z_{n}}+1}I_s
\approx I_s,
\end{aligned}
\end{equation}
From (\ref{eq_current_injection}), we see that almost all of the harmonic currents coming from node $n$ are being injected into the substation.
Accordingly, for every harmonic source we can define the \emph{harmonic current path} as the path of harmonic current which is flowing from the node with the harmonic source to the substation.
For example, the red line in Fig. \ref{fig_Sample_network} is a harmonic current path corresponding to the harmonic source at node $n$.
In a radial DN, there is only one harmonic current path for each harmonic source, but a branch of the {DN} can correspond to several harmonic current paths.
Thus, the harmonic currents of the branches falling on at least one harmonic current path can be the \emph{combination} of several harmonic sources.
For instance, if there are two harmonic sources with currents of 1 A and 2 A, then the possible combinations of the harmonic sources, i.e. the possible harmonic currents of the branches falling on at least one harmonic current path, could be 1 A, 2 A, and 3 A.

To identify the status of a branch as closed, it suffices to show that the branch falls on at least one harmonic current path.
Nevertheless, the status of a branch cannot be identified as either closed or open if the branch carries no harmonic current.
Rather, to identify the status of a branch as open one needs to {detect} a loop, in which only the single branch of interest does not fall on any harmonic current path.
In that case, the switch on the branch of interest must be open.

The above methodology in identifying open switches is based on the fact that, in a DN with radial topology, at least one branch {must} be open in every loop of branches.
For example, in the power system of Fig. \ref{fig_Sample_network} which includes only one loop {the branch $(q,q+1)$ must be open if} there is a harmonic current on every branch except on the branch $(q, q+1)$.

In summary, carrying no harmonic current is a necessary but not sufficient condition for a branch to be identified as open. The status of a switch with no harmonic current can be identified as open only by {detecting} a loop with the said conditions.
\section{Topology Identification in a Nonlinear Optimization Framework}\label{sec_MILP}
In this section, a TI scheme is proposed that utilizes both the fundamental and harmonic current measurements in DNs.
The proposed TI scheme is formulated as a nonlinear optimization problem.
\subsection{Partial Topology Identification using Harmonic Currents}\label{sec_MINLP}
In general, dedicated meters are installed in DNs to monitor the harmonic current injections of large nonlinear loads such as large DGs.
On this account, we assume {that} the harmonic current measurements are available for topology identification.
Let $I_{i}^h$ denote the current of the $h^{\rm th}$ harmonic component associated with the harmonic source located at node $i$.
According to the Kirchhoff's Current Law (KCL) we have:
\begin{equation}\label{eq_KCL-1}
\sum_{j\in \mathcal{N}_i} I^h_{(i,j)}=I^h_{i} \quad \forall k\in \mathcal{K},
\end{equation}
where $I^h_{(k,j)}$ is the $h^{\rm th}$ harmonic current on branch $(k,j)$; $\mathcal{K}$ is the set of nodes containing harmonic sources; and $\mathcal{N}_k$ is the set of nodes connected to the node $k$.
For the nodes with no harmonic source, we have:
\begin{equation}\label{eq_KCL-2}
\sum_{j\in \mathcal{N}_i} I^h_{(i,j)}=0 \quad \forall i\in \mathcal{N}, i\notin \mathcal{K},
\end{equation}
where $\mathcal{N}$ is the set of all nodes in the DN.

As was discussed in Section \ref{sec_analysis}, {detecting} the paths of harmonic currents is vital in identifying the status of the switches in the DN.
To this end, one needs to examine whether or not a branch carries a harmonic current. This can be done by looking into the numerical value of $I_{(i,j)}^h$.
More precisely, if $I^h_{(i,j)}$ is larger than the {magnitude of the smallest-magnitude} combination of all existing harmonic sources in the DN, {i.e. the parameter $z$,} then the branch must be on a {harmonic current} path.
To simplify the analysis, we {make two adjustments.
First,} instead of assessing the magnitude of harmonic current $I_{(i,j)}^h$ we choose to assess the absolute values of the real and imaginary parts of the harmonic current $I_{(i,j)}^h$.
{Second, instead of using the parameter $z$ in the comparisons, we use a threshold parameter $c$ that is less than the parameter $z$.}
To this end, we first obtain the absolute values of the real and imaginary parts of the harmonic current $I_{(i,j)}^h$ as follows:
\begin{equation}\label{eq_abs_re}
{\rm Re}\{I^h_{(i,j)}\}=2q^{h,r}_{(i,j)}X^{h,r}_{(i,j)}-X^{h,r}_{(i,j)} \quad \forall (i,j)\in \mathcal{B}
\end{equation}
\begin{equation}\label{eq_abs_im}
{\rm Im}\{I^h_{(i,j)}\}=2q^{h,i}_{(i,j)}X^{h,i}_{(i,j)}-X^{h,i}_{(i,j)} \quad \forall (i,j)\in \mathcal{B}
\end{equation}
\begin{equation}\label{eq_x}
X^{h,r}_{(i,j)}\geq 0;~X^{h,i}_{(i,j)}\geq 0 \quad\quad\quad\quad\quad\quad \forall (i,j)\in \mathcal{B},
\end{equation}
where $\mathcal{B}$ is the set of all branches; $q^{h,r}_{(i,j)}$ and $q^{h,i}_{(i,j)}$ are binary variables; and $X^{h,r}_{(i,j)}$ and $X^{h,i}_{(i,j)}$ are non-negative variables.
If the real or imaginary part of $I^h_{(i,j)}$ is positive, then $q^{h,r}_{(i,j)}$ or $q^{h,i}_{(i,j)}$ must be 1 and $X^{h,r}_{(i,j)}$ or $X^{h,i}_{(i,j)}$ must be equal to the real or imaginary part of $I^h_{(i,j)}$, respectively.
If the real or imaginary part of $I^h_{(i,j)}$ is negative, then $q^{h,r}$ or $q^{h,i}$ must be 0 and $X^{h,r}_{(i,j)}$ or $X^{h,i}_{(i,j)}$ must be equal to the absolute value of the real or imaginary part of $I^h_{(i,j)}$, respectively.
All in all, we see that the absolute values of the real and imaginary parts of $I^h_{(i,j)}$ are $X^{h,r}_{(i,j)}$ and $X^{h,i}_{(i,j)}$, respectively.

Next, we {compare} the summation of the variables $X^{h,r}_{(i,j)}$ and $X^{h,i}_{(i,j)}$ against the threshold $c$ to see whether or not the branch $(i,j)$ falls on any harmonic current path.
The harmonic current flowing through a branch may be the combination of several harmonic sources at different nodes of the DN.
If the branch $(i,j)$ is not on any harmonic current path, then we must have $X_{(i,j)}^{h,r}+X_{(i,j)}^{h,i}\leq c$.
In contrast, if the branch $(i,j)$ is on a harmonic current path, we must have $X_{(i,j)}^{h,r}+X_{(i,j)}^{h,i}\geq c$.
The comparison between $X_{(i,j)}^{h,r}+X_{(i,j)}^{h,i}$ and the threshold parameter $c$ can be mathematically formulated by the following constraint:
\begin{multline}\label{eq_path}
(1-b^h_{(i,j)})[(X^{h,r}_{(i,j)}+X^{h,i}_{(i,j)})-c]\\
+b^h_{(i,j)}[c-(X^{h,r}_{(i,j)}+X^{h,i}_{(i,j)})]\leq 0~~~~\forall (i,j)\in \mathcal{B},
\end{multline}
where $b^h_{(i,j)}$ is a binary variable.
From (\ref{eq_path}), the numerical values of 0 and 1 for the binary variable $b_{(i,j)}^h$ translate to the inequalities of $X_{(i,j)}^{h,r}+X_{(i,j)}^{h,i}\leq c$ and $X_{(i,j)}^{h,r}+X_{(i,j)}^{h,i} \geq c$, respectively.
Consequently, numerical values of $b_{(i,j)}^h=0$ and $b_{(i,j)}^h=1$ correspond to the branch $(i,j)$ not being on any harmonic current path and being on at least one harmonic current path, respectively.

From the discussion in section \ref{sec_analysis}, we know that a numerical value of $1$ for the binary variable $b_{(i,j)}^h$ testifies the in-service status of the branch $(i,j)$, but a numerical of $0$ for the binary variable $b_{(i,j)}^h$ doesn't indicate the in-service or out-of-service status for the branch.
As a result, additional information is needed to determine the status of all the branches.
This can be done by looking into the fundamental currents {measurements}.
\subsection{Topology Identification using Fundamental Currents}
According to the KCL we have:
\begin{equation}\label{eq_KCL-3}
\sum_{j\in N_i} I_{(i,j)}=I_{i} \quad \forall i\in \mathcal{N},
\end{equation}
where $I_{(i,j)}$ is the fundamental current on branch $(i,j)$; and $I_{i}$ is the fundamental current injection into the node $i$, which can be obtained from the pseudo-measurements, i.e., the measurements estimated from the seasonal load curves.
The status of the branch $(i,j)$ can be modeled by the following constraint:
\begin{equation}\label{eq_fun_lim}
  -Ms_{(i,j)}\leq I_{(i,j)} \leq Ms_{(i,j)} \quad \forall (i,j)\in \mathcal{B}.
\end{equation}
where $M$ is a large number selected arbitrarily and $s_{(i,j)}$ is a binary variable.
A numerical value of $1$ for the binary variable $s_{(i,j)}$ corresponds to the in-service status for the branch $(i,j)$ and requires the fundamental current $I_{(i,j)}$ to be within the interval of $[-M,M]$.
In contrast, a numerical value of $0$ for the binary variable $s_{(i,j)}$ corresponds to the out-of-service status for the branch $(i,j)$ and requires the fundamental current $I_{(i,j)}$ to be 0.
\subsection{Topology Identification using Fundamental and Harmonic Currents}
So far, we have explained how measurements of {fundamental and harmonic currents can be used to derive the status of the switches.}
However, to integrate both fundamental and harmonic into on TI problem, we need to derive the relationships between the binary variables $b^h_{(i,j)}$ and $s_{(i,j)}$.
$b^h_{(i,j)}=1$ indicates that the switch is closed which enforces $s_{(i,j)}=1$.
Also, $s^{(i,j)}=0$ indicates that the switch is open which enforces $b^h_{(i,j)}=0$.
Finally, when $b^h_{(i,j)}$ is 0 the corresponding switch can be open or closed and accordingly $s_{(i,j)}$ can be 0 or 1.
These relationships between the binary variables $b_{(i,j)}^h$ and $s_{(i,j)}$ can be modeled by the following constraint:
\begin{equation}\label{eq_b_s}
  b^h_{(i,j)}\leq s_{(i,j)} \quad \forall (i,j)\in \mathcal{B}.
\end{equation}

Besides the relationships between the fundamental and harmonic switch binary variables, there is a further relationship between the fundamental switch binary variables which should be taken into account.
The following equation establishes another relationship between the binary variables $s_{(i,j)}$, using the fact that in every loop of a radial network at least one switch must be open:
\begin{equation}\label{eq_loop}
  \sum_{(i,j)\in \mathcal{L}} s_{(i,j)}\leq N_{l}-1 \quad \forall \mathcal{L}\in \mathcal{P},
\end{equation}
where $\mathcal{L}$ is the set of all branches in an arbitrary loop of the DN; $\mathcal{P}$ is the set of all the possible loops in the DN, which can be formed using the algorithm provided in \cite{GabowFinding}; and $N_{l}$ is the number of branches in the loop $\mathcal{L}$.
For (\ref{eq_loop}), we notice that if in a loop $\mathcal{L}$ of the radial network $N_{l}-1$ switches turn out to be closed from the harmonic currents analysis, the binary variables $s_{(i,j)}$ for all these $N_l-1$ switches are enforced to be 1.
Then the binary variable $s_{(i,j)}$ for the last remaining switch of the loop $\mathcal{L}$ must be 0, which requires the last switch to be open.
Therefore, (\ref{eq_loop}) addresses the said condition at the end of section \ref{sec_analysis} to identify the status of the switch with no harmonic current.

Finally, there is another relationship between all the binary variables $s_{(i,j)}$ which ensures the radial configuration of the DN \cite{MIQP2016}:
\begin{equation}\label{eq_sum}
  \sum_{(i,j)\in \mathcal{B}} s_{(i,j)}= N-1,
\end{equation}
where $N$ is the total number of nodes in the DN.

Let $I^{h,m}_{(i,j)}$ and $I^{m}_{(i,j)}$ denote the measurements of harmonic and fundamental currents, respectively, collected by PMUs.
The following optimization problem minimizes the error in the estimation of harmonic and fundamental currents in the DN \cite{Mohammad2019}:
\begin{equation}\label{eq_MINLP}
\begin{aligned}
  &{\rm \textbf{Min}} \sum_{(i,j)\in \mathcal{M}} \frac{|I^h_{(i,j)}-I^{h,m}_{(i,j)}|}{|I^{h,m}_{(i,j)}|}+\frac{|I_{(i,j)}-I^{m}_{(i,j)}|}{|I^{m}_{(i,j)}|}\\
  &\textbf{s.t.}~~~~ {\rm Eqs}~(\ref{eq_KCL-1})\sim(\ref{eq_sum}),
\end{aligned}
\end{equation}
where $\mathcal{M}$ is the set of all branches equipped with PMUs.
It should be noted that, in problem (\ref{eq_MINLP}), we minimize the \emph{normalized} values of the errors.
If the accuracy is not the same for the fundamental and harmonic measurements, then we may include some coefficients in the objective function in (\ref{eq_MINLP}) to further adjust normalization.
\section{Topology Identification in a Linear Optimization Framework}
The proposed TI scheme formulated in (\ref{eq_MINLP}) is a mixed integer nonlinear programming (MINLP) {problem}, while no algorithm is yet developed to guarantee to solve this problem.
Thus, in this section we take several steps to reformulate the problem in (\ref{eq_MINLP}) to a solvable MILP format.

\subsection{Tackling the Non-Linearity in the Constraints}
In problem (\ref{eq_MINLP}), the constraints (\ref{eq_abs_re}), (\ref{eq_abs_im}), and (\ref{eq_path}) include non-linear terms. All the non-linear terms are formed by the product of a binary variable and a continuous variable.
{The non-linear  constraint (\ref{eq_abs_re}) in problem (\ref{eq_MINLP}) can be replaced with the following linear constraints:}
\begin{equation}\label{eq_abs_re1}
{\rm Re}\{I^h_{(i,j)}\}=2W^{h,r}_{(i,j)}-X^{h,r}_{(i,j)} \quad \forall (i,j)\in \mathcal{B}
\end{equation}
\begin{equation}\label{eq_abs_re3}
-M q^{h,r}_{(i,j)}\leq W^{h,r}_{(i,j)}\leq M q^{h,r}_{(i,j)}\quad \forall (i,j)\in \mathcal{B}
\end{equation}
\begin{equation}\label{eq_abs_re2}
-M(1-q^{h,r}_{(i,j)})\leq W^{h,r}_{(i,j)}-X^{h,r}_{(i,j)}\leq M(1-q^{h,r}_{(i,j)})\quad \forall (i,j)\in \mathcal{B},
\end{equation}
where {$W_{(i,j)}^{h,r}$} is a new continuous optimization variable.
A similar approach can be taken to replace the non-linear constraints (\ref{eq_abs_im}) and (\ref{eq_path}) with linear ones.
As a result, the constraint (\ref{eq_abs_im}) can be replaced by the following constraints:
\begin{equation}\label{eq_abs_im1}
{\rm Im}\{I^h_{(i,j)}\}=2W^{h,i}_{(i,j)}-X^{h,i}_{(i,j)} \quad \forall (i,j)\in \mathcal{B}
\end{equation}
\begin{equation}\label{eq_abs_im3}
-M q^{h,i}_{(i,j)}\leq W^{h,i}_{(i,j)}\leq M q^{h,i}_{(i,j)}\quad \forall (i,j)\in \mathcal{B}
\end{equation}
\begin{equation}\label{eq_abs_im2}
-M(1-q^{h,i}_{(i,j)})\leq W^{h,i}_{(i,j)}-X^{h,i}_{(i,j)}\leq M(1-q^{h,i}_{(i,j)})\quad \forall (i,j)\in \mathcal{B}.
\end{equation}
Similarly, the constraint (\ref{eq_path}) can be replaced by the following constraints:
\begin{equation}\label{eq_path1}
2cb^h_{(i,j)}-2X^{h}_{(i,j)}+X^{h,r}_{(i,j)}+X^{h,i}_{(i,j)}-c\leq 0 \quad \forall (i,j)\in \mathcal{B}
\end{equation}
\begin{equation}\label{eq_path3}
-M b^{h}_{(i,j)}\leq X^{h}_{(i,j)}\leq M b^{h}_{(i,j)}\quad\quad\quad\quad\quad\quad \forall (i,j)\in \mathcal{B}
\end{equation}
\begin{equation}\label{eq_path2}
\begin{aligned}
-M(1-b^{h}_{(i,j)})\leq X^{h}_{(i,j)}-(X^{h,i}_{(i,j)}+X^{h,i}_{(i,j)})\leq &M(1-b^{h}_{(i,j)})\\
&\forall (i,j)\in \mathcal{B}.
\end{aligned}
\end{equation}
\subsection{Tackling the Non-Linearity in the Objective Function}
The non-linearity of the objective function in problem (\ref{eq_MINLP}) is because of the optimization variables that are complex numbers and the operators that calculate the absolute values.
All these sources of non-linearity can be removed by adding the new variables $G_{(i,j)}^{h,r}$, $G_{(i,j)}^{h,i}$, $G_{(i,j)}^{r}$, and $G_{(i,j)}^{i}$ and the following new constraints {to the optimization problem (\ref{eq_MINLP})}:
\begin{equation}\label{eq_obj1}
 -G^{h,r}_{(i,j)}\leq \frac{{\rm Re} \{I^h_{(i,j)}-I^{h,m}_{(i,j)}\}}{|{\rm Re} \{I^{h,m}_{(i,j)}\}|}\leq G^{h,r}_{(i,j)}~\forall (i,j)\in \mathcal{M}
\end{equation}
\begin{equation}\label{eq_obj2}
 -G^{h,i}_{(i,j)}\leq \frac{{\rm Im} \{I^h_{(i,j)}-I^{h,m}_{(i,j)}\}}{|{\rm Im} \{I^{h,m}_{(i,j)}\}|}\leq G^{h,i}_{(i,j)}~\forall (i,j)\in \mathcal{M}
\end{equation}
\begin{equation}\label{eq_obj3}
 -G^{r}_{(i,j)}\leq \frac{{\rm Re} \{I_{(i,j)}-I^{m}_{(i,j)}\}}{|{\rm Re} \{I^{m}_{(i,j)}\}|}\leq G^{r}_{(i,j)}~\forall (i,j)\in \mathcal{M}
\end{equation}
\begin{equation}\label{eq_obj4}
 -G^{i}_{(i,j)}\leq \frac{{\rm Im} \{I_{(i,j)}-I^{m}_{(i,j)}\}}{|{\rm Im} \{I^{m}_{(i,j)}\}|}\leq G^{i}_{(i,j)}~\forall (i,j)\in \mathcal{M}.
\end{equation}
As a result, the problem in (\ref{eq_MINLP}) can be reformulated to the following MILP:
\begin{equation}\label{eq_MILP}
\begin{aligned}
  &{\rm \textbf{Min}} \sum_{(i,j)\in \mathcal{M}} G^{h,r}_{(i,j)}+G^{h,i}_{(i,j)}+G^{r}_{(i,j)}+G^{i}_{(i,j)}\\
  &\textbf{s.t.}~~~~ {\rm Eqs}~(\ref{eq_KCL-1})\sim(\ref{eq_KCL-2}),(\ref{eq_KCL-3})\sim(\ref{eq_sum}),(\ref{eq_abs_re1})\sim(\ref{eq_obj4}).
\end{aligned}
\end{equation}
\section{Observability Analysis}
This section provides an analysis on the number of and location of harmonic sources and PMUs that are needed to {ensure full observability of the DN.}
For simplifying the analysis we consider a TI scheme that utilizes only harmonic current measurements:
\begin{equation}\label{eq_MILP_harmonic}
\begin{aligned}
  &{\rm \textbf{Min}} \sum_{(i,j)\in \mathcal{M}} G^{h,r}_{(i,j)}+G^{h,i}_{(i,j)}\\
  &\textbf{s.t.}~~~~ {\rm Eqs}~(\ref{eq_KCL-1})\sim(\ref{eq_KCL-2}),(\ref{eq_b_s})\sim(\ref{eq_sum}),(\ref{eq_abs_re1})\sim(\ref{eq_obj4}).
\end{aligned}
\end{equation}
We define an \emph{independent} loop as a loop that doesn't include any other loop within. We have the following theorem:

\textbf{Theorem 1:} \emph{Consider a DN in which a PMU is placed in each independent loop. The status of all switches in this DN could be identified using only the harmonic current measurements, if there is a harmonic source at each node of the DN.}

\emph{\textbf{Proof:}}
\begin{itemize}
         \item Step 1: we show that under the said condition, all harmonic current paths can be detected. For a loop with $N_l$ nodes, one can formulate $N_l-1$ independent equations on the currents flowing in the $N_l-1$ branches of the loop, each equation corresponding to a known harmonic source that is measured at one of the $N_l$ nodes of the loop;  see \cite{Mohammad2019} for more details.
Another independent equation can be formulated for the measurement performed by the PMU that is placed in the loop.
In overall, the $N_l$ formulated equations will provide a \emph{unique} solution to the currents flowing in $N_l$ branches of the loop.
This solution is then used along with (\ref{eq_path}) to calculate numerical values of the binary variables $b_{(i,j)}^h$. Consequently, all harmonic current paths in the DN can be detected.
         \item Step 2: A branch that falls on a harmonic current path can be identified as closed. To complete the proof of the theorem, we only need to show that a branch must be open if it doesn't fall on any harmonic current path.  To this end, let $(n,n+1)$ denote a branch carrying no harmonic current.
There must be two different harmonic current paths from node $n$ to the substation and from node $n+1$ to the substation, to carry the {harmonic} currents {generated} by the harmonic sources at nodes $n$ and $n+1$.
The two harmonic current paths don't include the branch $(n,n+1)$ since it is assumed that no harmonic current flows through it.
The branches on the said two harmonic currents paths together with the branch $(n,n+1)$ make a loop, in which the branch $(n,n+1)$ is the only one with no harmonic current.
Therefore, from Section \ref{sec_analysis} the branch $(n,n+1)$ must be open to preserve the radial topology of the DN.
       \end{itemize}

\section{Performance Evaluation}
\begin{figure}
  \centering
  \includegraphics[width=9cm]{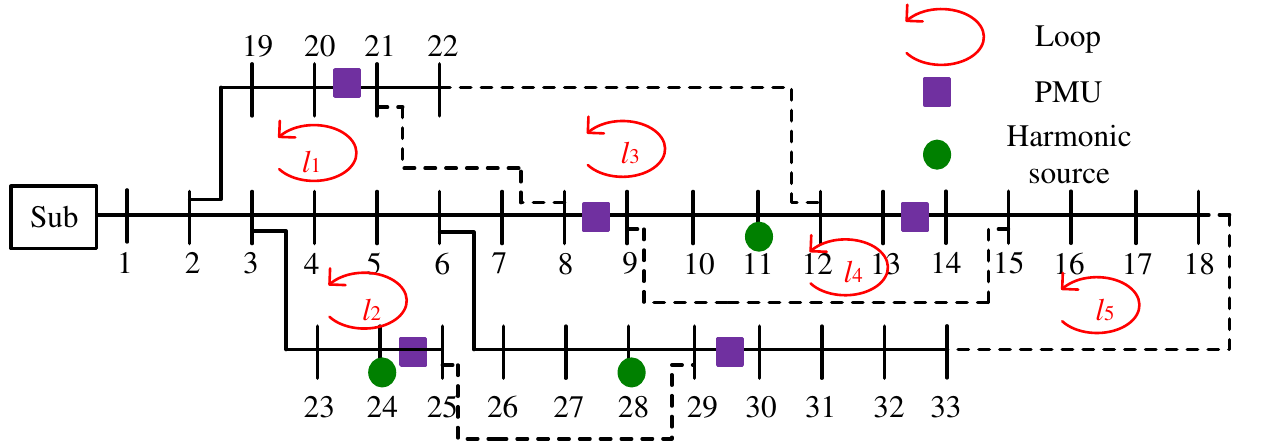}\\
  \caption{The IEEE 33-Bus power system includes five independent loops, where in each independent loop a PMU is placed. The dotted and dashed lines are normally open and closed lines, respectively.}\label{fig_33}
\end{figure}
In this section, IEEE 33-Bus power system \cite{33bus} is used for assessing the performance of the proposed TI scheme.
Unless stated otherwise, the magnitude and the location of harmonic source currents are as described in \cite{MeloHarmonic}, where there is data particularly on the 3$^{\rm rd}$ harmonic current of phase A.
This choice of data is following the fact that, the 3$^{\rm rd}$ harmonic currents normally come with the largest magnitudes.
There are 5 independent loops in the feeder under study as shown in Fig. \ref{fig_33}.
From \cite{Mohammad2019}, unique harmonic and fundamental branch current estimates can be obtained for the DN in Fig. \ref{fig_33}, which renders the problem (\ref{eq_MILP}) solvable.
Synthesized measurements of branch harmonic currents and harmonic sources are obtained from PSCAD, where the load is modeled based on the Model A given in \cite{Modeling}.
The measurements are contaminated according to the Gaussian distribution \cite{error}.
{Unless stated otherwise, the threshold parameter $c$ is set to 25$\%$ of the parameter $z$.
The results are obtained based on Monte Carlo simulation with 100 iterations for each test case.
The accuracy of the TI schemes can be obtained according to the following formula:
\begin{equation}\label{eq_accuracy}
  \text{Accuracy}=\frac{N_{\rm correct}}{N_{\rm total}}\times 100\%,
\end{equation}
where $N_{\rm total}$ is the total number of tests and $N_{\rm correct}$ is the number of the tests with a correct output being produced.
For a TI scheme that utilizes only the measurements of harmonic currents, i.e., the problem (\ref{eq_MILP_harmonic}), the correct output refers to the correct identification of harmonic current paths.
For the proposed scheme in (\ref{eq_MILP}) and the traditional TI scheme, the correct output refers to the correct topology of the DN.
{The topology of the DN resulted from the traditional TI scheme is obtained by solving the following optimization problem: }
\begin{equation}\label{eq_MILP_fundamental}
\begin{aligned}
  &{\rm \textbf{Min}} \sum_{(i,j)\in \mathcal{M}} G^{r}_{(i,j)}+G^{i}_{(i,j)}\\
  &\textbf{s.t.}~~~~ {\rm Eqs}~(\ref{eq_KCL-3})\sim(\ref{eq_fun_lim}),(\ref{eq_sum}).
\end{aligned}
\end{equation}

\subsection{The Overall Performances of the Proposed Scheme}

As discussed in section \ref{introduction}, the main source of error in the traditional TI scheme is the errors in pseudo-measurements.
The pseudo-measurements of fundamental current injections are estimated using the seasonal load curves and the nodal voltages \cite{Abur1995}.
Compared to the pseudo-measurements of the loads, larger amounts of error could be carried by the pseudo current injections.
When using only harmonic current measurements, the switches on the harmonic current paths are always identified correctly, which can be seen from the simulation results in Fig. \ref{fig_harmonic_current_path}.
Fig. \ref{fig_pseudo} shows the accuracy of the proposed and the traditional TI schemes for a various percentage of errors in the pseudo-measurements.
From Fig. \ref{fig_pseudo}, the accuracy of the proposed scheme is always higher than that of the traditional scheme.
As the percentage of error in pseudo-measurement increases, the accuracy of both schemes decreases.
When the percentage of error in pseudo-measurement becomes $90\%$, the proposed scheme outperforms the traditional scheme by a $13\%$ improvement in accuracy.
\begin{figure}
  \centering
  \includegraphics[width=9cm]{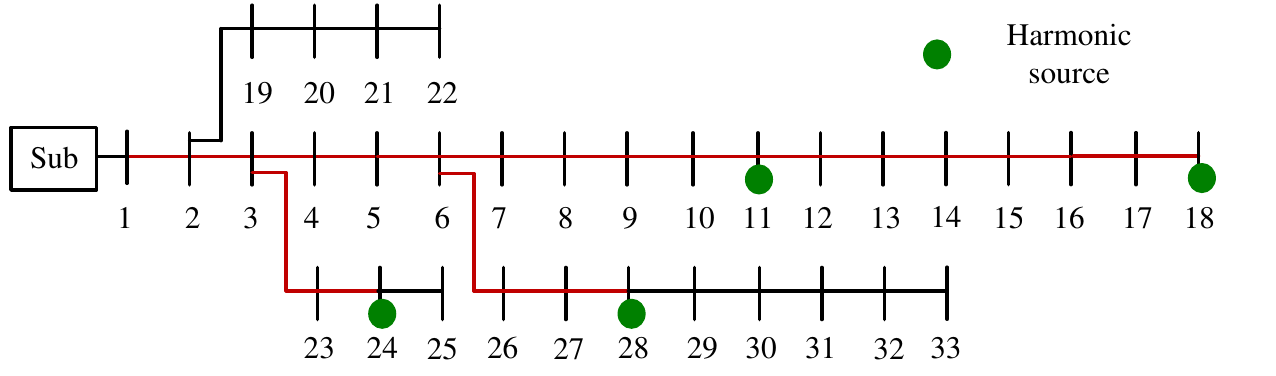}\\
  \caption{The harmonic current paths {detected by the TI scheme in (\ref{eq_MILP_harmonic})} are shown in red line.}\label{fig_harmonic_current_path}
\end{figure}

Fig. 6 also shows that, for 40$\%$ or less percentage of error in pseudo-measurements, the accuracy of the proposed and the traditional TI schemes are almost the same.
This result is justifiable by the fact that, the incorrect estimated status in the traditional TI scheme correspond to switches that are not on any harmonic current path and consequently cannot be corrected by the proposed TI scheme.
The better performance of the proposed TI scheme compared to the traditional scheme is seen {especially} when the percentage of error in pseudo-measurements becomes more than $40\%$.
\begin{figure}
  \centering
  \includegraphics[width=8cm]{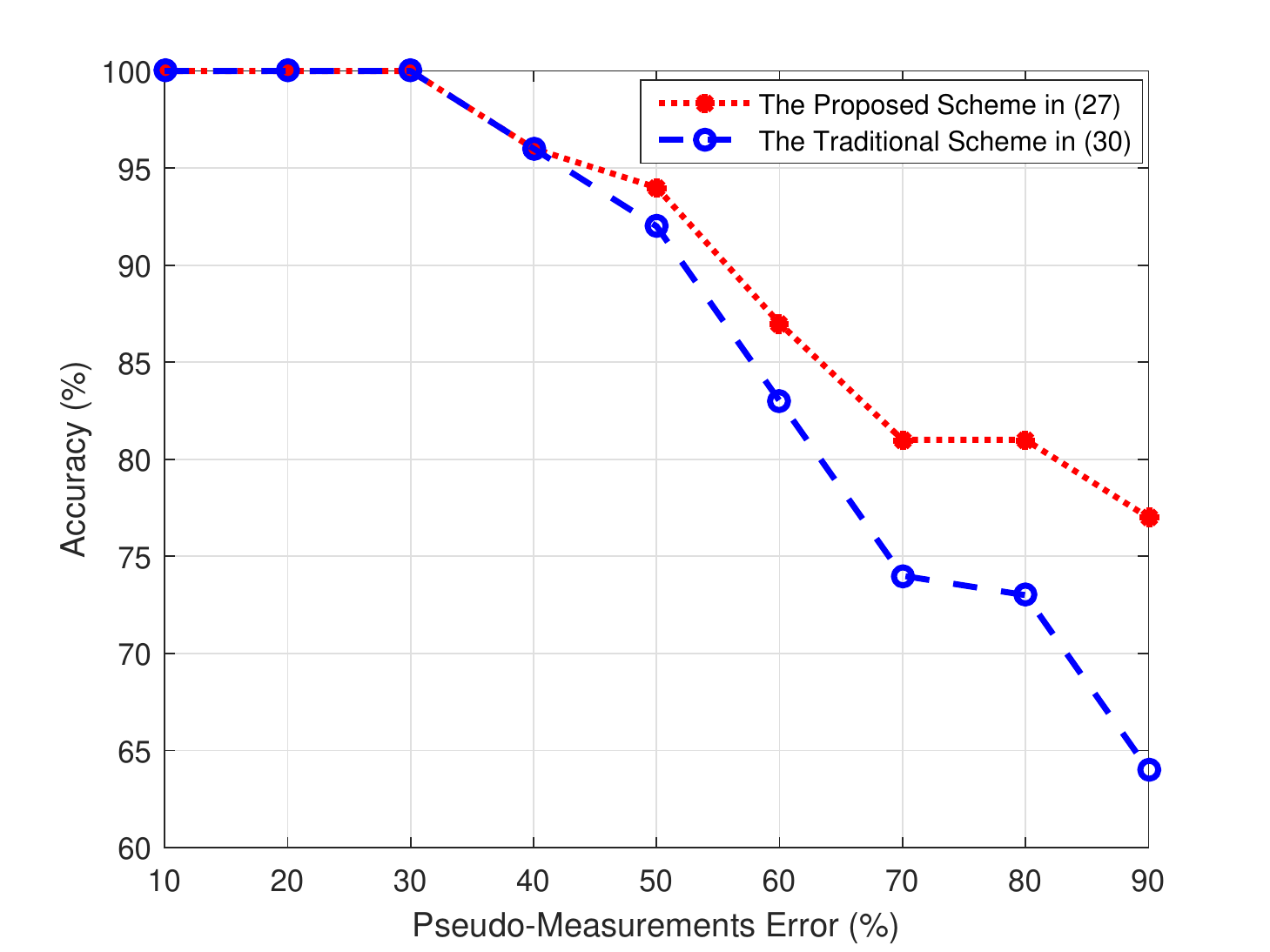}\\
  \caption{The accuracy of the proposed and the traditional TI schemes against errors in pseudo-measurements.}\label{fig_pseudo}
\end{figure}

\begin{figure}
  \centering
  \includegraphics[width=8cm]{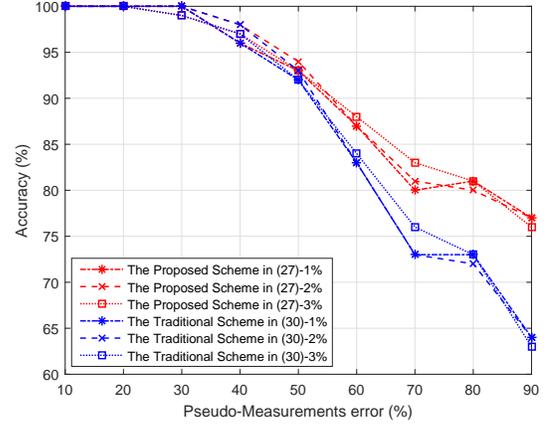}\\
  \caption{The accuracy of the proposed and the traditional TI schemes with joint fundamental measurement errors.}\label{fig_per_com}
\end{figure}


\subsection{The Impact of Erroneous Fundamental Current Measurements}
We repeat the same numerical case study discussed in Section VI-A with a minor change in the simulation setup, by considering erroneous PMUs measurements.
The range of total vector error (TVE) in the PMUs data is set to $1\%$ to $3\%$, following the IEEE Std. C37.118.1-2011 \cite{Power2011C37}.
For various percentage of pseudo-measurements, the accuracy of the proposed and the traditional TI schemes are almost identical to the case with errorless PMUs data.
This again shows the outperformance of the proposed TI scheme compared to the traditional TI scheme for the case of erroneous PMU measurements.


\subsection{The Impact of Erroneous Harmonic Current Measurements}\label{sec_test_harmonic}
This section assesses the impact of erroneous harmonic current measurements on the accuracy of the TI schemes.
We consider a scenario where there are $5\%$ TVE in the PMUs data, $90\%$ error in the pseudo-measurements, and $3\%$ error in the measurements of fundamental currents flowing through the branches.
From Fig. \ref{fig_harmonic_error_results}(a), for the case of erroneous harmonic current measurements all the branches on harmonic current paths can still be {detected} correctly.
Also, the proposed  TI scheme still comes with $13\%$ advantage in accuracy over the traditional TI scheme.
Therefore, it can be observed that the proposed TI scheme is robust to the errors in harmonic currents measurements.
\begin{figure}
  \centering
  \includegraphics[width=8cm]{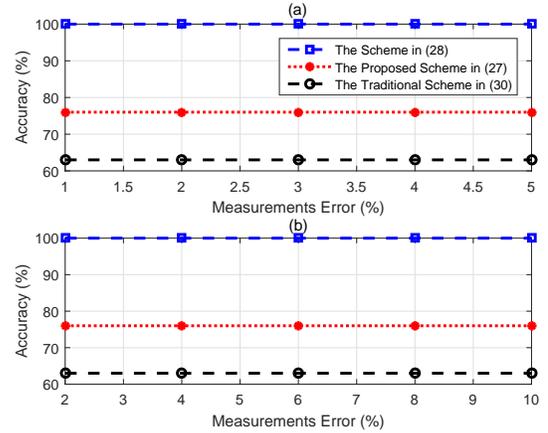}\\
  \caption{The accuracy of the TI schemes under (a) erroneous harmonic current measurements and (b) erroneous harmonic source measurements.}\label{fig_harmonic_error_results}
\end{figure}

To justify this observation, first, we notice that the value of the threshold is less than the value of the contaminated harmonic current measurement.
As a result, the harmonic current paths can be identified correctly.
Second, the problem (\ref{eq_MILP}) minimizes the total error in the estimation of harmonic and fundamental currents.
Since the harmonic currents are scattered all around the power system, incorrect identification of a harmonic current path is much more impactful on the objective function of problem (\ref{eq_MILP}), compared to incorrect identification of a switch status in the traditional TI scheme that uses fundamental current measurements.
For instance, the incorrect identification of the harmonic current path generated by the current source at node 18 may increase the objective function of problem (\ref{eq_MILP}) by more than $1$.
In contrast, incorrect identification of a switch status may increase the same objective function by only a few hundredth of $1$.
In conclusion, the optimization problem (\ref{eq_MILP}) tends to be robust to the errors in harmonic current measurements.

\subsection{The Impact of Erroneous Harmonic Source Measurements}

The measurements of harmonic sources are performed by the monitoring sensors that are located at the loads. These sensors may produce a level of error larger than the error being produced by PMUs.
Accordingly, in this section, we consider harmonic source measurements with up to $10\%$ error. The values of errors set in pseudo-measurements and fundamental current measurements are $90\%$ and $3\%$, respectively.
Fig. \ref{fig_harmonic_error_results}(b) shows the accuracy of the TI schemes for a various percentage of harmonic source TVE. From Fig. \ref{fig_harmonic_error_results}(b), the proposed TI scheme is robust to the errors in harmonic source measurements. This observation is justifiable by a reasoning similar to the one provided in Section \ref{sec_test_harmonic}.


\subsection{The Impact of the Threshold Parameter}
The threshold parameter $c$ should be set in a way that, {in the proposed TI scheme} the impact of errors in harmonic sources and harmonic currents are lowered.
In this section, the threshold parameter is set to various levels from $5$ to $25$ percent of the parameter $z$ {defined in section \ref{sec_MINLP}}.
The errors in harmonic currents, harmonic sources, pseudo-measurements, and fundamental currents are $5\%$, $10\%$, $90\%$ and $3\%$, respectively.

\begin{figure}
  \centering
  \includegraphics[width=8cm]{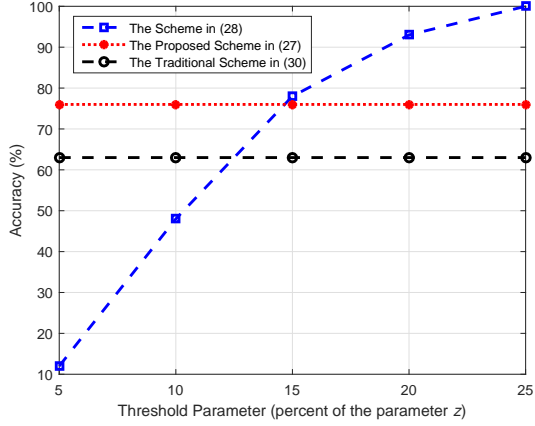}\\
  \caption{The impact of threshold parameter on TI accuracy ($\%$).}\label{fig_threshold2}
\end{figure}

\begin{figure}
  \centering
  \includegraphics[width=8.5cm]{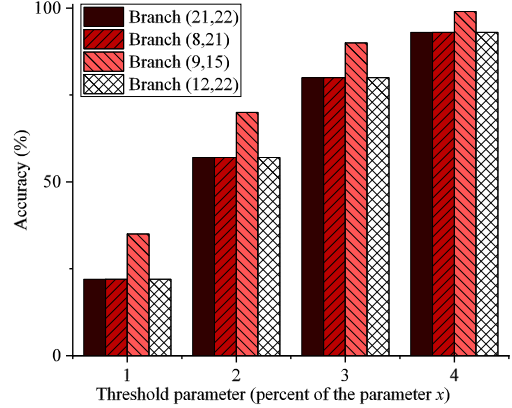}\\
  \caption{Accuracy of the TI scheme in (\ref{eq_MILP_harmonic}) for various values of the threshold parameter. The results are shown only for switches with incorrect identified status.}\label{fig_threshold}
\end{figure}
Fig. \ref{fig_threshold2} and Fig. \ref{fig_threshold} show the accuracy of the TI schemes for various values of the threshold parameter $c$.
A low value for the threshold parameter, e.g. $c=5\%z$ may lead to incorrect detection of some harmonic current paths.
For the TI scheme that uses only harmonic current measurements, this can result in incorrect identification of some branch status, e.g. {the status of} branches (8,21) and (9,15), as it can be seen from Fig. \ref{fig_threshold}.
In contrast, Fig. \ref{fig_threshold2} shows that the accuracy of the proposed TI scheme is the same for various values of the threshold parameter $c$.
This is due to the fact that, the proposed TI scheme in (\ref{eq_MILP_harmonic}) utilizes the fundamental current measurements to identify the switch status correctly, even if the harmonic current paths are identified incorrectly.
Consequently, we can observe that the proposed TI scheme is robust to the choice of the threshold parameter $c$ because of using both the harmonic and fundamental measurements together.

\begin{figure}
  \centering
  \includegraphics[width=8cm]{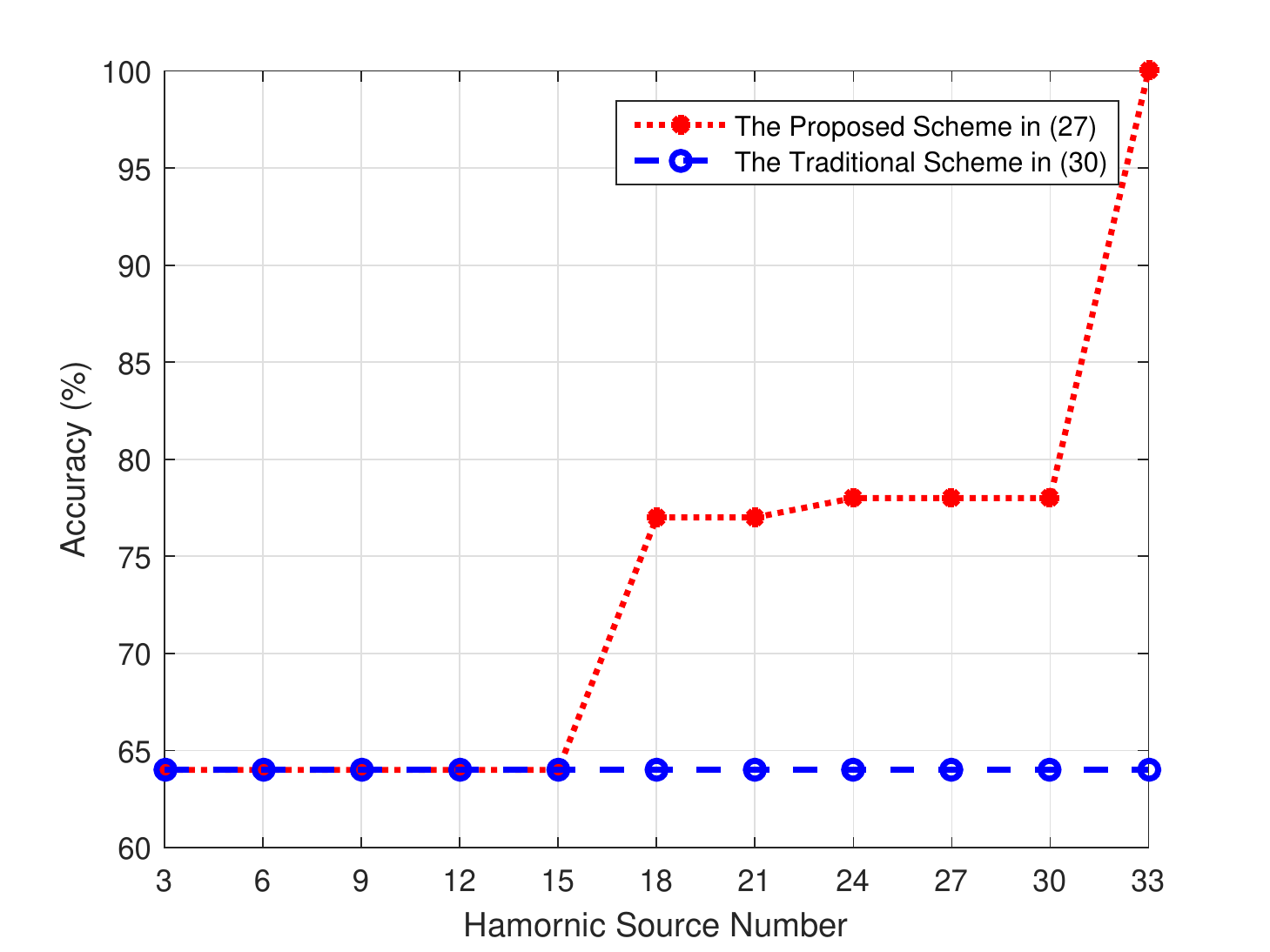}\\
  \caption{The impact of the number of harmonic source on TI accuracy ($\%$).}\label{fig_theorem}
\end{figure}
\subsection{The Impact of the Number of Harmonic Sources}

In this section, we study the impacts of the number of harmonic sources on TI accuracy.
Here, we still assume that five PMUs are installed in our test system as shown in Fig. \ref{fig_33}.
It is assumed that when there are $n$ harmonic sources in the test network, then these harmonic sources are located at node 1 to node $n$.
The harmonic source current at each node is the same as the harmonic source current at node 11 which is shown in Fig. \ref{fig_33}.
It is assumed that there are no harmonic measurement errors, and the threshold parameter $c$ is set to 10$\%$ of the parameter $z$.
The errors in pseudo-measurements are set to $90\%$.

The accuracy of the TI schemes for different number of harmonic sources are shown in Fig. \ref{fig_theorem}.
As the number of harmonic sources increases, the accuracy of the proposed TI scheme increases.
This is because by having more harmonic sources, more switches fall on harmonic current paths, which help the proposed TI scheme to correctly derive the status of the switches that would be identified incorrectly if one uses only the fundamental current measurements.
Also, the proposed TI scheme works with $100\%$ accuracy when the number of harmonic sources is equal to the number of nodes in the DN, which is explained as a requirement for full observability in theorem 1.

There is a chance that the proposed TI scheme works with $100\%$ accuracy even with fewer sources.
For instance, when there are harmonic sources at nodes 18, 22, 25, and 33, all the closed switches fall on at least one harmonic current path, and the proposed TI scheme also achieves $100\%$ accuracy.
However, the theorem gives a sufficient condition to make sure that all closed switches fall on at least one harmonic current path, which consequently leads to 100$\%$ accuracy.
\section{Conclusion}

{In this paper, a novel topology identification (TI) scheme was proposed for radial distribution networks (DNs).
The proposed TI scheme uses both the harmonic and fundamental current measurements to enhance the TI accuracy in DNs.
Through numerical simulations on IEEE 33-Bus power system, it was shown that the accuracy of the proposed TI scheme is at least $10\%$ more than that of the traditional TI scheme that uses only measurements of fundamental currents.
Also, it was shown that the proposed TI scheme is robust to the errors in harmonic and fundamental current measurements.
Furthermore, a theoretical analysis was provided on the observability of the DN using harmonic current measurements.}
Accordingly, a case study is shown that the proposed TI scheme will work with $100\%$ accuracy when the number of harmonic sources is equal to the number of nodes in the DN.


\ifCLASSOPTIONcaptionsoff
  \newpage
\fi

\bibliographystyle{IEEEtran}
\bibliography{MILP}

\begin{thebibliography}{10}
\providecommand{\url}[1]{#1}
\csname url@samestyle\endcsname
\providecommand{\newblock}{\relax}
\providecommand{\bibinfo}[2]{#2}
\providecommand{\BIBentrySTDinterwordspacing}{\spaceskip=0pt\relax}
\providecommand{\BIBentryALTinterwordstretchfactor}{4}
\providecommand{\BIBentryALTinterwordspacing}{\spaceskip=\fontdimen2\font plus
\BIBentryALTinterwordstretchfactor\fontdimen3\font minus
  \fontdimen4\font\relax}
\providecommand{\BIBforeignlanguage}[2]{{%
\expandafter\ifx\csname l@#1\endcsname\relax
\typeout{** WARNING: IEEEtran.bst: No hyphenation pattern has been}%
\typeout{** loaded for the language `#1'. Using the pattern for}%
\typeout{** the default language instead.}%
\else
\language=\csname l@#1\endcsname
\fi
#2}}
\providecommand{\BIBdecl}{\relax}
\BIBdecl

\bibitem{Mohammad2018}
M.~{Farajollahi}, A.~{Shahsavari}, E.~M. {Stewart}, and H.~{Mohsenian-Rad},
  ``Locating the source of events in power distribution systems using
  micro-{PMU} data,'' \emph{IEEE Trans. Power Syst.}, vol.~33, no.~6, pp.
  6343--6354, Nov 2018.

\bibitem{DSSE2017}
A.~{Primadianto} and C.~{Lu}, ``A review on distribution system state
  estimation,'' \emph{IEEE Trans. Power Syst.}, vol.~32, no.~5, pp. 3875--3883,
  Sep 2017.

\bibitem{Paolo2019}
P.~A. {Pegoraro}, K.~{Brady}, P.~{Castello}, C.~{Muscas}, and A.~{von Meier},
  ``Line impedance estimation based on synchrophasor measurements for power
  distribution systems,'' \emph{IEEE Trans. Instrum. Meas.}, vol.~68, no.~4,
  pp. 1002--1013, Apr 2019.

\bibitem{HM1999}
L.~{Mili}, G.~{Steeno}, F.~{Dobraca}, and D.~{French}, ``A robust estimation
  method for topology error identification,'' \emph{IEEE Trans. Power Syst.},
  vol.~14, no.~4, pp. 1469--1476, Nov 1999.

\bibitem{GSE1998}
K.~A. {Clements} and A.~S. {Costa}, ``Topology error identification using
  normalized lagrange multipliers,'' \emph{IEEE Trans. Power Syst.}, vol.~13,
  no.~2, pp. 347--353, May 1998.

\bibitem{GSE2004}
E.~M. {Lourenco}, A.~S. {Costa}, and K.~A. {Clements}, ``Bayesian-based
  hypothesis testing for topology error identification in generalized state
  estimation,'' \emph{IEEE Trans. Power Syst.}, vol.~19, no.~2, pp. 1206--1215,
  May 2004.

\bibitem{GSE2015}
E.~M. {Lourenço}, E.~P.~R. {Coelho}, and B.~C. {Pal}, ``Topology error and bad
  data processing in generalized state estimation,'' \emph{IEEE Trans. Power
  Syst.}, vol.~30, no.~6, pp. 3190--3200, Nov 2015.

\bibitem{Joint2012}
V.~{Kekatos} and G.~B. {Giannakis}, ``Joint power system state estimation and
  breaker status identification,'' in \emph{Proc. of North American Power
  Symposium}, Champaign, IL, USA, 2012.

\bibitem{Mohammad2019}
M.~{Farajollahi}, A.~{Shahsavari}, and H.~{Mohsenian-Rad}, ``Topology
  identification in distribution systems using line current sensors: An milp
  approach,'' \emph{IEEE Trans. Smart Grid}, pp. 1--1, Aug 2019.

\bibitem{Joint2020}
A.~{Gandluru}, S.~{Poudel}, and A.~{Dubey}, ``Joint estimation of operational
  topology and outages for unbalanced power distribution systems,'' \emph{IEEE
  Trans. Power Syst.}, pp. 1--1, Aug 2019.

\bibitem{MIQP2016}
Z.~{Tian}, W.~{Wu}, and B.~{Zhang}, ``A mixed integer quadratic programming
  model for topology identification in distribution network,'' \emph{IEEE
  Trans. Power Syst.}, vol.~31, no.~1, pp. 823--824, Jan 2016.

\bibitem{Abur1995}
A.~{Abur}, D.~{Shirmohammadi}, and C.~S. {Cheng}, ``Estimation of switch
  statuses for radial power distribution systems,'' in \emph{Proc. of IEEE Int.
  Symp. Circuits and Syst.}, vol.~2, Seattle, WA, USA, 1995.

\bibitem{GPS2009}
A.~Carta, N.~Locci, and C.~Muscas, ``{GPS}-based system for the measurement of
  synchronized harmonic phasors,'' \emph{IEEE Trans. Instrum. Meas.}, vol.~58,
  no.~3, pp. 586--593, Mar 2009.

\bibitem{Carlo2009}
A.~Carta, N.~Locci, and C.~Muscas, ``A {PMU} for the measurement of
  synchronized harmonic phasors in three-phase distribution networks,''
  \emph{IEEE Trans. Instrum. Meas.}, vol.~58, no.~10, pp. 3723--3730, Oct 2009.

\bibitem{Japan2003}
H.~{Ukai}, K.~{Nakamura}, and N.~{Matsui}, ``Dsp- and gps-based synchronized
  measurement system of harmonics in wide-area distribution system,''
  \emph{IEEE Trans. Ind. Electron.}, vol.~50, no.~6, pp. 1159--1164, Dec 2003.

\bibitem{Lei2018}
L.~Chen, W.~Zhao, Q.~Wang, F.~Wang, and S.~Huang, ``Dynamic harmonic
  synchrophasor estimator based on sinc interpolation functions,'' \emph{IEEE
  Trans. Instrum. Meas.}, pp. 1--12, Nov 2018.

\bibitem{HPE2019}
L.~{Chen}, W.~{Zhao}, F.~{Wang}, and S.~{Huang}, ``Harmonic phasor estimator
  for p class phasor measurement units,'' \emph{IEEE Trans. Instrum. Meas.},
  pp. 1--1, May 2019.

\bibitem{Extended2014}
M.~Chakir, I.~Kamwa, and H.~L. Huy, ``Extended {C}37.118.1 {PMU} algorithms for
  joint tracking of fundamental and harmonic phasors in stressed power systems
  and microgrids,'' \emph{IEEE Trans. Power Del.}, vol.~29, no.~3, pp.
  1465--1480, Jun 2014.

\bibitem{MeloHarmonic}
I.~D. Melo, J.~L. Pereira, A.~M. Variz, and P.~A. Garcia, ``Harmonic state
  estimation for distribution networks using phasor measurement units,''
  \emph{Electr. Power Syst. Res.}, vol. 147, pp. 133--144, Jun 2017.

\bibitem{MELO2019303}
I.~D. Melo, J.~L. Pereira, P.~F. Ribeiro, A.~M. Variz, and B.~C. Oliveira,
  ``Harmonic state estimation for distribution systems based on optimization
  models considering daily load profiles,'' \emph{Electr. Power Syst. Res.},
  vol. 170, pp. 303 -- 316, May 2019.

\bibitem{Hamed2017}
M.~Farajollahi, A.~Shahsavari, and H.~Mohsenian-Rad, ``Location identification
  of high impedance faults using synchronized harmonic phasors,'' in
  \emph{Proc. of IEEE PES Innov. Smart Grid Technol. Conf.}, Washington, DC,
  USA, 2017.

\bibitem{DSS}
D.~Montenegro, R.~Dugan, and G.~Ramos, ``Harmonics analysis using
  sequential-time simulation for addressing smart grid challenges,'' in
  \emph{Proc. of 23rd Int. Conf. Electricity Distribution}, Lyon, France, 2015.

\bibitem{Akagi2000}
H.~{Fujita}, T.~{Yamasaki}, and H.~{Akagi}, ``A hybrid active filter for
  damping of harmonic resonance in industrial power systems,'' \emph{IEEE
  Trans. Power Electron。}, vol.~15, no.~2, pp. 215--222, Mar 2000.

\bibitem{GabowFinding}
H.~N. Gabow and E.~W. Myers, ``Finding all spanning trees of directed and
  undirected graphs,'' \emph{Siam Journal on Computing}, vol.~7, no.~3, pp.
  280--287, Aug 1978.

\bibitem{33bus}
M.~E. {Baran} and F.~F. {Wu}, ``Network reconfiguration in distribution systems
  for loss reduction and load balancing,'' \emph{IEEE Trans. Power Del.},
  vol.~4, no.~2, pp. 1401--1407, Apr 1989.

\bibitem{Modeling}
T.~F. on~Harmonics~Modeling and Simulation, ``Modeling and simulation of the
  propagation of harmonics in electric power networks. i. concepts, models, and
  simulation techniques,'' \emph{IEEE Trans. Power Del.}, vol.~11, no.~1, pp.
  452--465, Jan 1996.

\bibitem{error}
R.~{Singh}, B.~C. {Pal}, and R.~B. {Vinter}, ``Measurement placement in
  distribution system state estimation,'' \emph{IEEE Trans. Power Syst.},
  vol.~24, no.~2, pp. 668--675, May 2009.

\bibitem{Power2011C37}
\emph{I{EEE} Standard for Synchrophasor Measurements for Power Systems}.\hskip
  1em plus 0.5em minus 0.4em\relax IEEE Std C37.118.1-2011 (Revision of IEEE
  Std C37.118-2005), Dec 2011.

\end{thebibliography}

\end{document}